# Observation of a crossover in kinetic aggregation of Palladium colloids.

M. Ghafari[1], M. Ranjbar[2] and S. Rouhani[1]

[1]Department of Physics, Sharif University of Technology, Tehran PO Box 11135-9161, Iran.
[2]Department of Physics, Isfahan University of Technology, Isfahan PO Box 84156-83111, Iran

**Abstract**

We use field emission scanning electron microscope (FE-SEM) to investigate the growth of palladium colloids over the surface of thin films of $WO_3$/glass. The film is prepared by Pulsed Laser Deposition (PLD) at different temperatures. A $PdCl_2$ (aq) droplet is injected on the surface and in the presence of steam hydrogen the droplet is dried through a reduction reaction process. Two distinct aggregation regimes of palladium colloids are observed over the substrates. We argue that the change in aggregation dynamics emerges when the measured water drop Contact Angel (CA) for the $WO_3$/glass thin films passes a certain threshold value, namely $CA \approx 46°$, where a crossover in kinetic aggregation of palladium colloids occurs. Our results suggest that the mass fractal dimension of palladium aggregates follows a power-law behavior. The fractal dimension $(D_f)$ in the fast aggregation regime, where the measured CA values vary from 27° up to 46° according to different substrate deposition temperatures, is $D_f = 1.75(\pm 0.02)$ – The value of $D_f$ is in excellent agreement with kinetic aggregation of other colloidal systems in fast aggregation regime. Whereas for the slow aggregation regime, with $CA = 58°$, the fractal dimension changes abruptly to $D_f = 1.92(\pm 0.03)$. We have also used a modified Box-Counting method to calculate fractal dimension of gray-level images and observe that the crossover at around $CA \approx 46°$ remains unchanged.

**Introduction**

The aggregation or flocculation of small particles to form larger structures and clusters have attracted serious attention, since a broad spectrum of aggregation processes including gelation and polymerization processes in polymer science [1], coagulation processes in aerosol and colloidal physics, percolation and nucleation in phase transitions and critical phenomena, rouleaux formation by red blood cell adhesion in hematology and crystallization and dendritic growth processes in material science can be well-described by scaling concepts [2,3]. In particular, several experimental investigations have been carried out for the irreversible kinetic aggregations in aqueous colloidal systems [4–7] suggesting that the resulting clusters are fractals. These investigations were followed by intense theoretical and computational activities to elucidate their statistical properties [8–10].

In a recent study [11], $WO_3$/glass thin films were prepared by pulsed laser deposition (PLD) at 100 mTorr oxygen pressure. In the presence of $H_2$, a wet-gasochromic switching with an edge-to-center coloring nature was observed when aqueous $PdCl_2$ was used as hydrogen catalyst. Hydrogen gas acts as a common reducing agent with no residual chemical impact on the system; hence, the resulting catalyst layer has high purity. Palladium nanoparticles were formed over the substrate through a reduction reaction inside the droplet and a peculiar aggregation of Pd colloids on a continuous gray layer of palladium was reported.

Colloidal nanoparticles mostly enter structured materials as one-, two-, or three-dimensional networks, grown at a solid-solid, fluid-fluid, or solid-fluid interface. Colloidal particles are solid or liquid matters with various sizes ranging from a few nanometers up to many micrometers which are suspended inside a gas, liquid, or solid. These particles demonstrate essential properties compared to small molecules, because of their mesoscopic size with the resulting large surface-to-volume ratio and intermediate dynamics [12]. Unlike highly concentrated suspensions that are close to their gelation point at drop deposition, the drying of low to semi-concentrated nanocolloids form open aggregates of lower surface coverage $\Gamma_s$ and fractal dimension $D_f \leq d_{space}$. Such growth processes are critically determined by interpaticle interactions, particle-substrate interactions, and drying kinetics [13]. One of the forms of these colloids are solid particles suspended in a liquid medium, such as noble metal catalysts (Pd, Pt) inside an aqueous, which is what we are referring to here.

We have found that the resulting clusters possess different fractal dimensions in accordance with different contact angle values in two regimes, namely the fast aggregation regime with $CA \lesssim 46°$, and the slow aggregation regime with $CA \gtrsim 46°$, which governs the kinetics of aggregation in each regime. This work is the first direct attempt to relate the hydrophobicity of the substrate to the fractal properties of the colloidal clusters formed over it.

This paper is organized as follows. In the first section the details of the experimental setup is provided. In the second section, a method to interpret the fractal properties of aggregates using their two-dimensional projected image is introduced, the validity of the measurements is argued, and the possibility of having three dimensional aggregates is offered. Section three mainly elaborates this issue by implementing a modified Box-Counting method to calculate the fractal dimension of gray-level images and indicates that the results achieved in section two remains unchanged. Finally we discuss the implications of our work.

## 1. Experimental Setup

Thin $WO_3$ films were fabricated by pulsed laser deposition (PLD) method from tungsten oxide pressed powder onto circular glass substrates of 11 mm diameter, which prior to deposition, were cleaned ultrasonically by methanol and DI water. The employed deposition system was a stainless steel chamber with a base pressure of $1 \times 10^{-5}$ Torr. In order to deposit stoichiometric tungsten oxide films, the PLD process was carried out in 100 mTorr oxygen (purity 99.9%) pressure. To obtain films with different surface hydrophilicities, the substrates were held at different temperatures during deposition process by means of an electric heater directly placed on the back of substrates. The processes of deposition were done under different substrate temperatures (STs) including room temperature, 100, 200, 300 and 400 °C. To ablate the tungsten oxide targets, 5000 pulses of a KrF laser (λ= 248 nm, τ =10 ns, laser energy of 200 mJ and R.R = 10 Hz) was delivered to the surface of the rotating target at a 45° angle. The substrate to target distance was kept constant at 7 cm. In order to make the palladium precursor , a $PdCl_2$ solution were prepared by solving 0.02 g $PdCl_2$ powder

(5N) into 99.9 cm³ DI water and 0.1 cm³ HCl. Then drops with constant volumes of 0.07 cm³ of the 0.2 g/l PdCl₂ solution were put on the surface of WO₃ layers. Finally, the palladium nano-particles were synthesized inside the drops by hydrogen-reduction method. To do this, a constant flow (2 l/min) of 10% $H_2$/Ar mixture gas was delivered into a sealed chamber contacting $PdCl_2$/$WO_3$/glass samples.

The surface morphology of obtained $PdCl_2$/$WO_3$/glass samples was observed on a FE-SEM (Hitachi model S4460) instrument. Hydrophilicity of $WO_3$/glass samples were investigated by the contact angle measurements which were performed in atmospheric air at room temperature using a commercial contact angle meter (Data physics OCA 15plus) with ±1° accuracy (see Figure 1). A droplet was injected on the surface using a 2 μl micro-injector.

Figure 2 shows a set of FE-SEM images (scale bars=5 μm) from aggregation of Pd nanoparticles on $WO_3$/glass substrates with different surface hydrophobicity in the drop-drying process. Because of the metallic nature of palladium nano-particles, their brightness is much higher than the non-metallic $WO_3$ substrates in electron microscopy and could be easily recognized in our FE-SEM images from the flat substrates. The aggregates resemble to fractal likes structures of different shapes and branching depending on the hydrophobicity of $WO_3$ substrate.

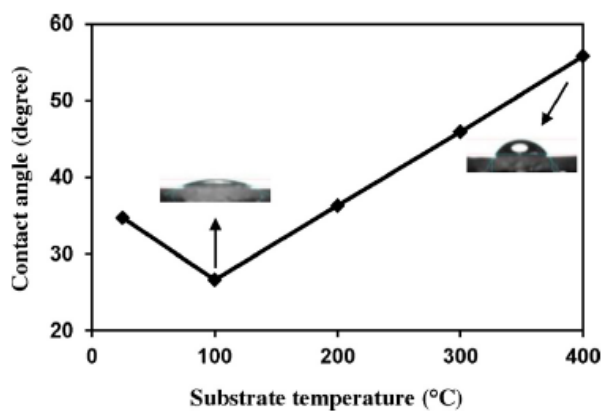

Figure 1. Contact angle (CA) of water for the WO3/glass thin films as a function of substrate temperature.

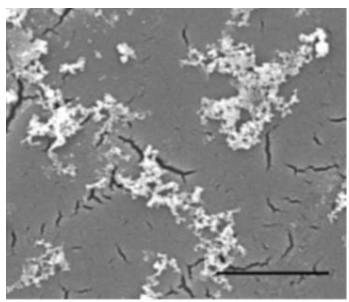 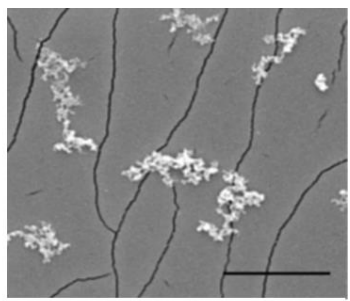

(a)  
CA=33 ° (ST=25 °C)

(b)  
CA=27 ° (ST=100 °C)

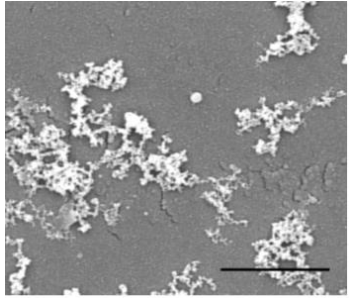

**(c)**

CA=36 ° (ST=200 °C)

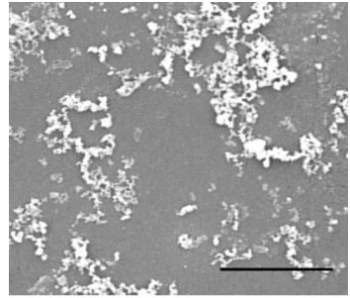

**(d)**

CA=46 ° (ST=300 °C)

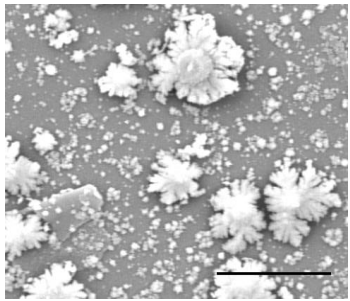

**(e)**

CA=58 ° (ST=400 °C)

*Figure 2. Shows a set of FE-SEM images of Pd/WO3 (scale bars= 5 μm). The contact angle of water (CA measurements) as hydrophobicity factor is shown at each substrate deposition temperature from (a) to (e).*

## 2. Interpretation of the Fractal Properties from Two-dimensional Projected Image: Results and Discussion

Since we are examining the FE-SEM images, our analysis is performed for a two dimensional projection of the clusters. Although the aggregates are three-dimensional inside the solution, we assume that upon drying through reduction reaction they collapse to form nearly flat, two-dimensional structures and we essentially tried to extract their 2-d mass fractal dimension to see whether it shows scale invariance properties or not. There are several methods to find the mass fractal dimension of a structure from its two dimensional projected image, namely the Nested Square Method (NSM), Perimeter Grid Method (PGM), and ensemble method (EM) [14]. Here we apply the Nester Square Method mainly because we have a few ensemble of clusters – five clusters at each substrate deposition temperature. For calculating the fractal dimension of clusters using NSM, we first turn the images into binary sequences, then partition them into squares of increasing size from 1×1 square pixels up to 40×40 square pixels – note that the size of each cluster is roughly 160×160 square pixels – and finally count the number ($N$) of squares needed to cover the entire cluster. This method is basically similar to the ordinary Minkowski-Bouligand dimension in which we count the number of boxes required to cover a set:

$$dim_{box}(S) = \lim_{\epsilon \to 0} \frac{\ln(N)}{\ln(1/\epsilon)} \tag{1}$$

The 2-d mass fractal dimension ($D_f$) is estimated by the linear regression slope of the $Ln - Ln$ plot of the $1/boundary$ against pixel numbers ($N$) (Figure 3).

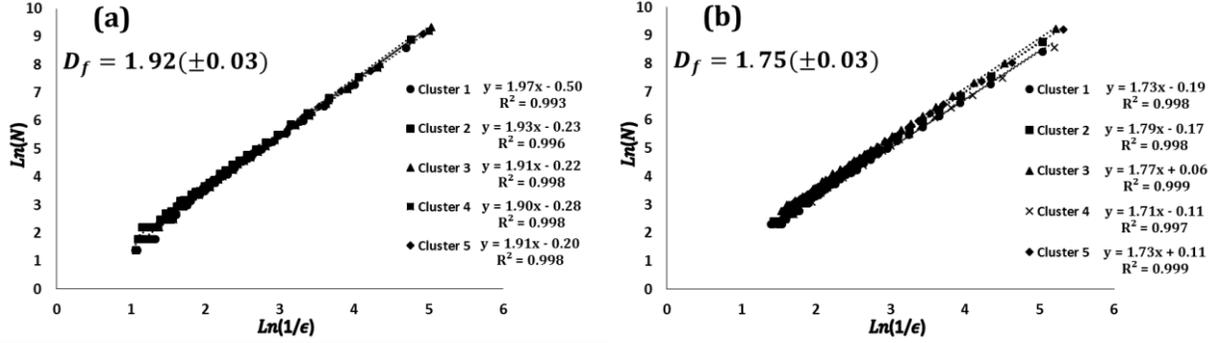

Figure 3. Plot of Ln (N) vs. Ln (1/ϵ), where the dashed lines are the least-square fit to the data of each cluster. Approximate fractal dimension is the average over an ensemble of five clusters with CAs equal to (a) 58 °, and (b) 46 °. Similar scale-invariance trend can be seen for clusters with CA values 36 °, 33 °, and 27 ° with measured fractal dimension 1.74(±0.03), 1.71(±0.05), and 1.77(±0.03), respectively. A slight variation in the y-intercept of each regression line is believed to happen due to different size magnifications of the images and therefore caused by the resolution of each image.

The plots indicate that the fractal dimension follows a power-law behavior for each of the clusters and they can be indeed treated as two-dimensional fractals. It can also be inferred that there is a crossover in the fractal dimension of aggregates. For aggregates with $CA \lesssim 46°$ the average fractal dimension is $D_f = 1.75(\pm 0.02)$, whereas for the dense aggregates with $CA = 58°$ the fractal dimension jumps to $D_f = 1.92(\pm 0.03)$. We calculated the average fractal dimension for the fast aggregation regime, with $D_f = 1.75(\pm 0.02)$, using the weighted average method (see Equation 2). $D_{f_i} = x_i \pm \sigma_i (for\ i = 1, 2, 3, \wedge 4)$ represents the estimate fractal dimension of an ensemble of aggregates for $CA$ values 46°, 36°, 33°, and 27°, respectively. To calculate the weighted average we have:

$$D_{f_{average}} = \frac{\sum_{i=1}^{4} \frac{x_i}{\sigma_i^2}}{\sum_{i=1}^{4} \frac{1}{\sigma_i^2}}$$

$$\sigma_{average} = \left(\sum_{i=1}^{4} \frac{1}{\sigma_i^2}\right)^{\frac{-1}{2}} \tag{2}$$

According to the experiment in ref [11], the rate of gasochromic switching and consequently the formation palladium aggregates on the surface decreases considerably when $CA = 58°$– coloration process gradually happens at about 17-30 minutes for the fast aggregation regime (at 300 °C, 200 °C, 100 °C, and 25 °C substrate deposition temperatures) and 51 minutes for the slow aggregation regime (at 400 °C). This, in fact, suggests that the hydrophobicity of the substrate is a major factor responsible for controlling the rate of deposition as well as the aggregation of the particles. These particles originate from hydrogen-induced nucleation inside the solution and are inherently under random Brownian motion. We believe that as the hydrophilicity decreases, the droplet has less tendency to wet the surface and as a results the concentration of the Pd colloids inside the aqueous will increase. Since it takes much longer for the nanoparticles to aggregate at lower hydrophilicity, the particles have more time to diffuse inside the solution and attract each other to form a more dense branching fractal-like shape with a fractal dimension value closer to d=2.

We then attempted to see if the internal structure of clusters exhibits scale invariance behavior and to further investigate if any phase transition occurs near $CA \approx 46°$ by finding the two-point correlation density function $c(r)$ using the equation below:

$$c(r) = \frac{\Gamma\left(\frac{d}{2}\right)}{2\pi^{d/2}\Delta r} \langle \frac{M(r+\Delta r) - M(r)}{r^{d-1}} \rangle \tag{3}$$

Where $d$ is the dimension of space, $r$ is the location of a particle and $\Delta r$ is the distance between two particles in the pixel co-ordinate system, $\langle ... \rangle$ represents the averaging over the position of particles, and $M(r)$ is equal to 1 at the location of a particle and 0 elsewhere. For a two-dimensional projection of these fractals, if we take $l$ as the monomer diameter in the pixel co-ordinates, we expect $c(r) \propto r^{-\alpha}$ for $l \leq r \ll R_g$ where $\alpha = 2 - D_f$ and $R_g$ is the radius of gyration ($R_g$ varies from 40 to 60 pixels for relatively different cluster sizes) and can be estimated in the pixel co-ordinates using the following equation [15]:

$$R_g^2 = \frac{1}{N_{pixels}} \sum_{i=1}^{N_{pixels}} (r_i - r_{mean})^2 \tag{4}$$

Where $r_{mean}$ is the location of centre of mass and $N_{pixels}$ is the number of counted pixels of the projected image of each cluster.

However, due to the finite size effects of the clusters we are not able to precisely show that $c(r)$ exhibits a long range power-law behavior. It can be clearly seen from Figure 2 that the clusters have many regions of overlapping particles – especially at 400 °C deposition temperature where clusters with giant two-dimensional spherical cores in the middle of the aggregates exist – and therefore the curves are noisy (see Figure 4). Yet, an extended linear regime of correlation function can still be recognized in the range of $r \approx 5$ to $r \approx 40$ in the pixel coordinate system.

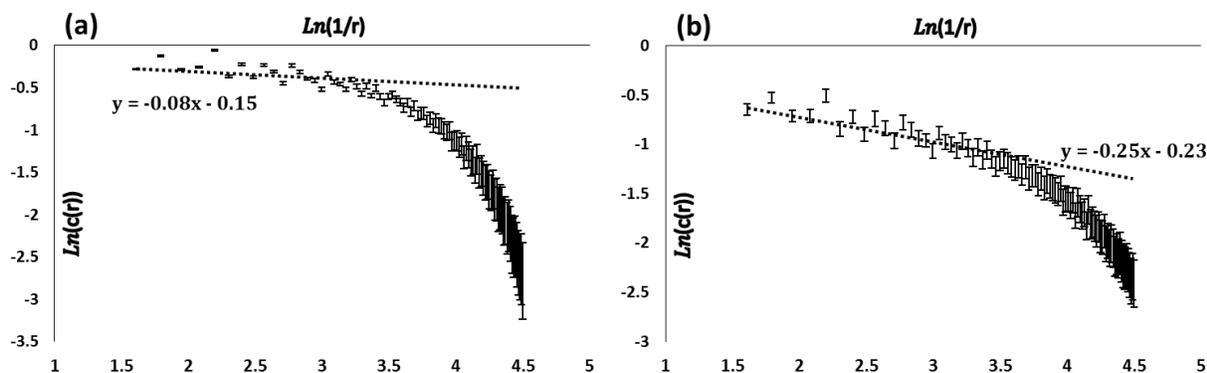

*Figure 4. Plot of Ln(c(r)) vs. Ln(r), where the dashed lines are the expected slope of the regression line according to the measured mass fractal dimension ($\alpha = 2 - D_f$). Each graph is the average two-point density correlation function of five clusters. (a) Approximate correlation density function for clusters with $CA = 58°$; the average radius of gyration is 40 pixels, so we expect to see the power-law behaviour in the range $1.61 \leq ln(r) < 3.69$. (b) Approximate correlation density function for clusters with $CA = 46°$; the average radius of gyration is 53 pixels, so we expect to see the power-law behaviour in the range $1.61 \leq ln(r) < 3.97$; same trends are seen for clusters with CA values 36°, 33°, and 27°.*

The limiting slopes, shown by dashed lines in Figure 4 suggest that the set of data points for two-point correlation function are consistent with the measured Hausdorff dimension $D_f = 2 - \alpha$ from the NSM method. Similar measurements on the two-point correlation functions and fractal dimension of gold and silica colloidal aggregates with $D_f$ 1.75 for fast aggregation and $D_f$ 2.05 for the slow aggregation regime was also reported elsewhere (see ref [4,5,16,17]) which reasonably coincide with the results of our experimental data. Therefore, we tried to investigate the possibility of having a phase transition by plotting the scaling exponent $\alpha$ as a function of contact angle (CA) and observe a drop in the value of $\alpha$ for $CA = 58°$:

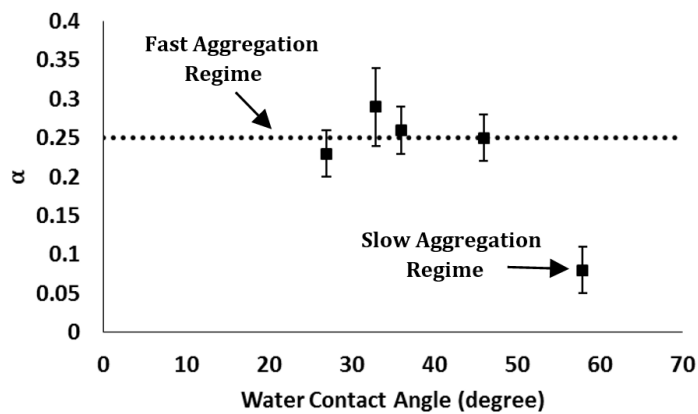

*Figure 5. Plot of α vs. water contact angle (CA). The dashed line is the predicted value of α in the fast aggregation regime. Each point represents the value of α exponent averaged over an ensemble of five clusters which changes with respect to different contact angle measures. (CAs vary from 27° to 58°).*

Universality in colloid aggregation has been checked for three chemically different colloidal systems under diffusion-limited and reaction-limited aggregation conditions [18]. Here, we offer what might be another candidate in colloidal systems which exhibits universal behavior in the limiting cases of fast and slow aggregation conditions. However, more precise measurements on the aggregation process is needed to reach this conclusion – especially near $CA \approx 46°$.

Strictly speaking, we do not offer any proof of assuming that the collapse of the aggregates is geometrical under projection from three to two dimensions. However, using an efficient method in image processing (described below); we are able to show that although the value of 3D fractal dimension is different from the projected two-dimensional images, the crossover point still exists.

## 3. A Modified Box-Counting Method to Calculate Fractal Dimension of Gray-Level Images.

Fractal geometry has gradually established its significance in the study of image characteristics when Pentland [19,20] provided the first theory about the human perception of smoothness and roughness of surfaces, with fractal dimension of 2 corresponding to smooth surfaces and fractal dimension of 3 corresponding to a maximum rough surface and considered the image intensity surface as fractal Brownian function (fBf) and measured the fractal dimension from Fourier power spectrum of fBf. Since then, many other theories such as reticular cell counting [21] and variations of box-counting method [22–24], which were applicable to a wider class of fractals, were introduced. Amongst these different approaches was a successful method introduced by Sarkar and Chaudhuri, also known as the differential box-counting (DBC) method, which proves to possess a relatively fast algorithm and yield more accurate results compared to the previous methods [25]. The DBC method is introduced as follows: First, consider an image of size $M \times M$ as a three-dimensional spatial surface with $(x, y)$ denoting 2-d position and $z$ denoting the gray level of the projected image – the gray level increases in value as the image surface intensity increases. Then, the $(x, y)$ space is partitioned into non-overlapping grids of size $s \times s$, such that $M/2 \geq s > 1$ where $s$ is an integer. Similar to the ordinary box-counting method (see Equation 1), in order to count the total number of pixels $(N_s)$ in the scale $s$ using differential box-counting method, we use the following procedure. On each grid there is a column of boxes of size $s \times s \times s'$, where $s'$ indicates the height of each box and is defined such that $\lfloor G/s' \rfloor = \lfloor M/s \rfloor$, where $\lfloor \ldots \rfloor$ denotes the floor function and $G$ is the total number of gray levels. For instance, you can see the measured number of boxes for an arbitrary image intensity in Figure 6 (where $s = 5$, and $s' = 4$). Let the minimum and maximum gray level in the $(i, j)$ grid fall into $k$th and $l$th boxes, respectively. The boxes covering this grid are counted in number as:

$$n_s(i,j) = l - k + 1 \tag{5}$$

Where $n_s(i,j)$ is the contribution of $N_s$ in $(i,j)$ grid and the subscript $s$ denotes the scale factor. In Figure 6, for example, we have $n_s(i,j) = 4 - 1 + 1 = 4$.

Considering contributions from all grids, then $N_s$ can be measured:

$$N_s = \sum_{i,j} n_s(i,j) \tag{6}$$

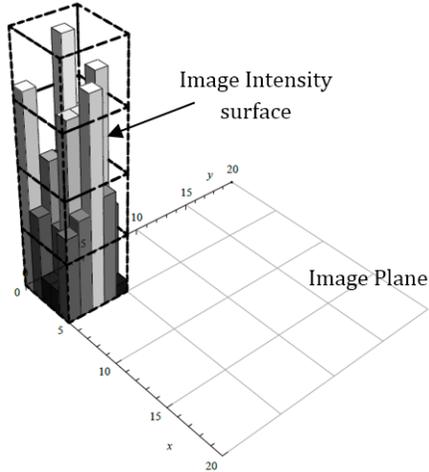

*Figure 6. Determination of Ns using DBC method, where the dashed cubes are the number of boxes needed to cover the grid (1, 1) and the gray bars represent the image intensity surface – each bar represents the gray level value of one pixel in the the image plane.*

Finally, the fractal dimension can be estimated by measuring the slope of the straight line best fitting the points in *Ln ($N_s$) vs. Ln (1/s)* plot [25]. However, this classical DBC method has some drawbacks – extensively discussed in [26,27] – such as over-counting or under-counting the number of boxes. Here, we briefly address one of the major drawbacks and then take a modified DBC method [26], known as the Shifting Differential Box-counting (SDBC) algorithm, so as to improve the accuracy of estimating the fractal dimension.

The DBC method can potentially over-count the number of boxes covering the image intensity surface. For an instance, as it can be seen from Figure 7, if the boxes are appropriately shifted along the z direction, no more than three boxes are needed to cover the gray level variation of the intensity surface on a specific block of the image plane.

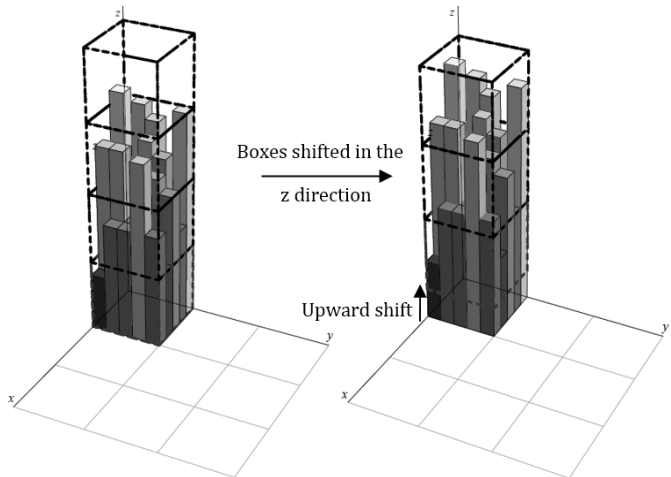

*Figure 7. If the boxes are shifted in the z direction (see the arrow in the right image), the number of boxes needed to count the intensity surface will be reduced from 4 to 3.*

As mentioned earlier in this section, the Differential Box-Counting method subdivides the gray level variation into some intervals of size *s* and measures the number of boxes required to cover the image intensity in the *z* direction by computing $\lceil z_{max}/s \rceil - \lfloor z_{min}/s \rfloor$, where $\lceil \dots \rceil$ denotes the ceiling function. In the Shifting Differential Box-counting algorithm, we measure the number of boxes by computing $\lceil \frac{z_{max}-z_{min}+1}{s} \rceil$. In this way, we resolve the problem of quantization effect, discussed above, and count the least number of boxes required for covering the image intensity surface more effectively. It has been proven that the SDBC algorithm obtains the estimate fractal dimension closer to the exact value compared to the DBC method [26]. Hence, by applying the SDBC algorithm to our FE-SEM images, we have found a sudden change in fractal dimension from $D_f = 2.56(\pm 0.01)$ in the fast aggregation regime to $D_f = 2.47(\pm 0.06)$ in the slow aggregation regime. The reason for emphasizing that there is a *sudden* change in fractal dimension is mainly because the fractal dimension does not vary noticeably in the fast aggregation regime; the fractal dimension is $D_f = 2.64(\pm 0.03)$ for $CA = 27°$, $D_f = 2.65(\pm 0.03)$ for $CA = 36°$, and $D_f = 2.60(\pm 0.05)$ for $CA = 46°$ where it drops to $D_f = 2.47(\pm 0.06)$ for $CA = 58°$ (see Figure 8).

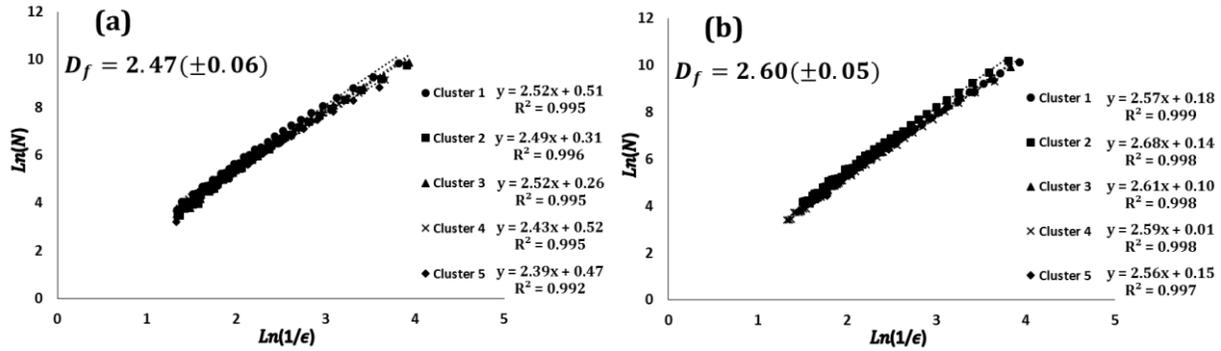

Figure 8. The plot of Ln (Ns) vs. Ln (1/ϵ) using the SDBC method, where $\epsilon = s/M$ and the dashed lines are the least-square fit to the data of each cluster. Approximate fractal dimension is the average over ensemble of five clusters with CAs equal to (a) 58 °, and (b) 46 °. Similar scale-invariance trend can be seen for clusters with CA values 36 °, 33 °, and 27 ° with measured fractal dimension 2.65($\pm$0.03), 2.54($\pm$0.01), and 2.64($\pm$0.03), respectively. A slight variation in the y-intercept of each regression line is believed to happen due to different size magnifications of the images and therefore caused by the resolution of each image.

It is also worth pointing out that if we take a closer look at the clusters at different substrate temperatures (see Figure 2), it is evident that the clusters at 400°C are fairly smoother (lower variation in gray intensity surface) than any other cluster at different substrate temperatures.

This result indicates that even by treating each cluster as a three-dimensional aggregate, the crossover can still be observed and this is a clear signal that geometric nature of the objects is changing when $CA \approx 46°$.

### 4. Summary and Discussion

In this paper we have discussed the observation of an abrupt change in fractal dimension of DLA-like clusters formed when depositing Palladium on WO3/glass substrate. We argue that the crossover happens when the CA value passes a certain threshold, namely $CA \approx 46°$. The fractal dimension of the aggregates changes from $D_f = 1.75(\pm 0.03)$ to $D_f = 1.92(\pm 0.03)$. Since the measured fractal dimension of palladium colloid aggregates happens to be similar to earlier studies on gold, silica, and polystyrene colloid aggregates, the possibility of having two limiting regimes of colloidal aggregation is also discussed. Abrupt changes in the geometry of aggregates have been observed under various conditions (for example see [28–32]). Most of these changes happen when temperature is the agent. Here we believe that the direct agent is hydrophobicity. To our knowledge this is the first time that hydrophobicity has been found to be the agent of crossover. Also in most of the reported geometrical changes, in fact self-similarity disappears altogether, whereas here we observe that the scaling behavior remains intact.